\documentclass[twocolumn,aps,10pt,prl,showpacs,showkeys,]{revtex4}
\usepackage[dvips]{graphicx}

\begin{document}
\title{Magnetic properties and the electronic structure of LiCoO$_{2}$}
\author{R. J. Radwanski}
\affiliation{Center of Solid State Physics, S$^{nt}$Filip 5,
31-150 Krakow, Poland\\
Institute of Physics, Pedagogical University, 30-084 Krakow,
Poland}
\author{Z. Ropka}
\affiliation{Center of Solid State Physics, S$^{nt}$Filip 5,
31-150 Krakow, Poland} \homepage{http://www.css-physics.edu.pl}
\email{sfradwan@cyf-kr.edu.pl}

\begin{abstract}
We have described properties of LiCoO$_{2}$ within the Quantum
Atomistic Solid State (QUASST) theory taking into account very
strong electron correlations, predominantly of the intra-atomic
origin, spin-orbit coupling and the detailed local
crystallographic surroundings and its symmetry. Properties of
LiCoO$_{2}$ are consistently explained together with NaCoO$_{2}$
and LaCoO$_{3}$ - in all of these compounds the Co$^{3+}$ ions
occur in the low-spin state. This low-spin state is the effect of
the relatively strong crystal-field interactions ($B_{4}^{z}$ =
+320 K $\Longleftrightarrow$ $10Dq$ = 3.3 eV) and is a
manifestation of the very large orbital moment of the Co$^{3+}$
ion.

\pacs{71.10.-w; 75.10.Dg } \keywords{electronic structure, crystal
field, spin-orbit coupling, LiCoO$_{2}$}
\end{abstract}

\maketitle \vspace {-0.3 cm}
\section{Introduction}\vspace {-0.2 cm}
Magnetic properties of LiCoO$_{2}$ are intriguing by at least 30
years. It is an insulator with a wide gap of 2.7 eV and exhibits
non-magnetic ground state \cite{1}. Such a non-magnetic ground
state is a surprise owing to strong magnetism in metallic
elemental Co, a ferromagnet with $T_{c}$ of 1395 K, and an ionic
oxide CoO, antiferromagnet with $T_{N}$ of 292 K. On other side
there exists another well-known oxide LaCoO$_{3}$ which exhibits
a non-magnetic ground state. LaCoO$_{3}$ has been studied by us
over last years getting consistent description of its magnetic
properties and electronic structure \cite{2} within the Quantum
Atomistic Solid State (QUASST) theory \cite{3} taking into
account very strong electron correlations and the effect of local
surroundings by means of the crystal-field parameters. In similar
approach we have described properties of antiferromagnet CoO
\cite{4}.

A scientific interest to LiCoO$_{2}$ has recently grown up due to
its potential applications in thermopower devices \cite{5} and
due to discovery of superconductivity in a sister-compound
NaCoO$_{2}$, when hydrated and off-stoichiometral
Na$_{x}$CoO$_{2}$$\cdot$yH$_{2}$O \cite {6,7}. Theoretical, called
"first principles", studies of properties of LiCoO$_{2}$ within
the band picture have been presented in Refs \cite {1,8} whereas
for NaCoO$_{2}$ in Ref. \cite{9}.

An aim of this paper is two fold. At first, we would like to
explain magnetic properties of LiCoO$_{2}$, in particular its
nonmagnetic ground state within the localized $d$-electron
picture, following the early works of Van Vleck, and secondly we
would like to determine its electronic structure, which can be
experimentally verified. It turns out that the physics of
magnetic properties and of the electronic structure is much
similar to that found in LaCoO$_{3}$, but with a much
deeper-in-energy non-magnetic singlet state and an opposite
trigonal distortion.\vspace {-0.5 cm}

\section{Theoretical outline}
\vspace {-0.2 cm} We describe properties of LiCoO$_{2}$ within the
Quantum Atomistic Solid State (QUASST) theory \cite{3,10,11},
which starts the description of a 3d-4f-5f compound from analysis
of the electronic structure of the involved ions. In QUASST we
take into account already from the beginning very strong electron
correlations, predominantly of the intra-atomic origin,
spin-orbit coupling and the detailed local crystallographic
surroundings and its symmetry. The insulating ground state of
LiCoO$_{2}$ we understand as resulting from the static ionic
charge distribution Li$^{1+}$Co$^{3+}$O$_{2}^{2-}$ established
during the formation of the compound. Magnetic properties of
LiCoO$_{2}$ we attribute to the Co$^{3+}$ ions because Li$^{1+}$
and O$^{2-}$ ions are magnetically inactive having closed
electron shells. Thus the main point of description of
LiCoO$_{2}$ within the QUASST theory is the very detailed
description of the electronic structure of the Co$^{3+}$ ions in
the crystallographic structure of LiCoO$_{2}$.

LiCoO$_{2}$ has a rhombohedral structure, which belongs to the
space group R$\bar{3}$m ($D_{3d}^{5}$) \cite{1,9}. The unit cell,
with parameters a = 496 pm and $\alpha$ = 32$^{o}$58$^{'}$,
contains only one chemical formula unit, and the Co, Li, and O
atoms occupy the 1a(0,0,0), 1b(1/2,1/2,1/2) and 2c(x,x,x) with
x=0.24, respectively. The rhombohedral structure of LiCoO$_{2}$
can be viewed as the one derived from the $NaCl$ rocksalt
structure of the pure CoO where every second plane of Co atoms
stacked in the main diagonal direction is replaced by a plane of
Li atoms. Thus, LiCoO$_{2}$ can be regarded as the layered
structure - neutron diffraction results confirm the fully ordered
state. For comparison, the rocksalt CoO structure with the
conventional lattice parameter $a_{c}$ = 426 pm can be described
as the rhombohedral lattice with $a_{r}$ = 521.7 pm and $\alpha$
= 33$^{o}$33$^{'}$. The Co atoms in CoO occupy 1a and 1b sites of
the R$\bar{3}$m space group whereas 2c site with x=0.25
(exactly!) is occupied by oxygen atoms.

The above presented crystallographic analysis reveals that the
cobalt ion in LiCoO$_{2}$ compound is placed in a local,
slightly-distorted, oxygen octahedron. Its diagonal is along the
rhombohedral axis. In such structural construction a trigonal
distortion can be easily realized by elongation or compression of
the distance between the whole oxygen planes. Exactly the same
construction of the O-Co-O planes occurs in the perovskite
structure, in LaCoO$_{3}$ \cite {2} for instance, perpendicularly
to the cube diagonal, and even in FeBr$_{2}$ \cite {12}, having a
hexagonal unit cell.

\begin{figure}
\begin{center}
\includegraphics[width = 10.0 cm]{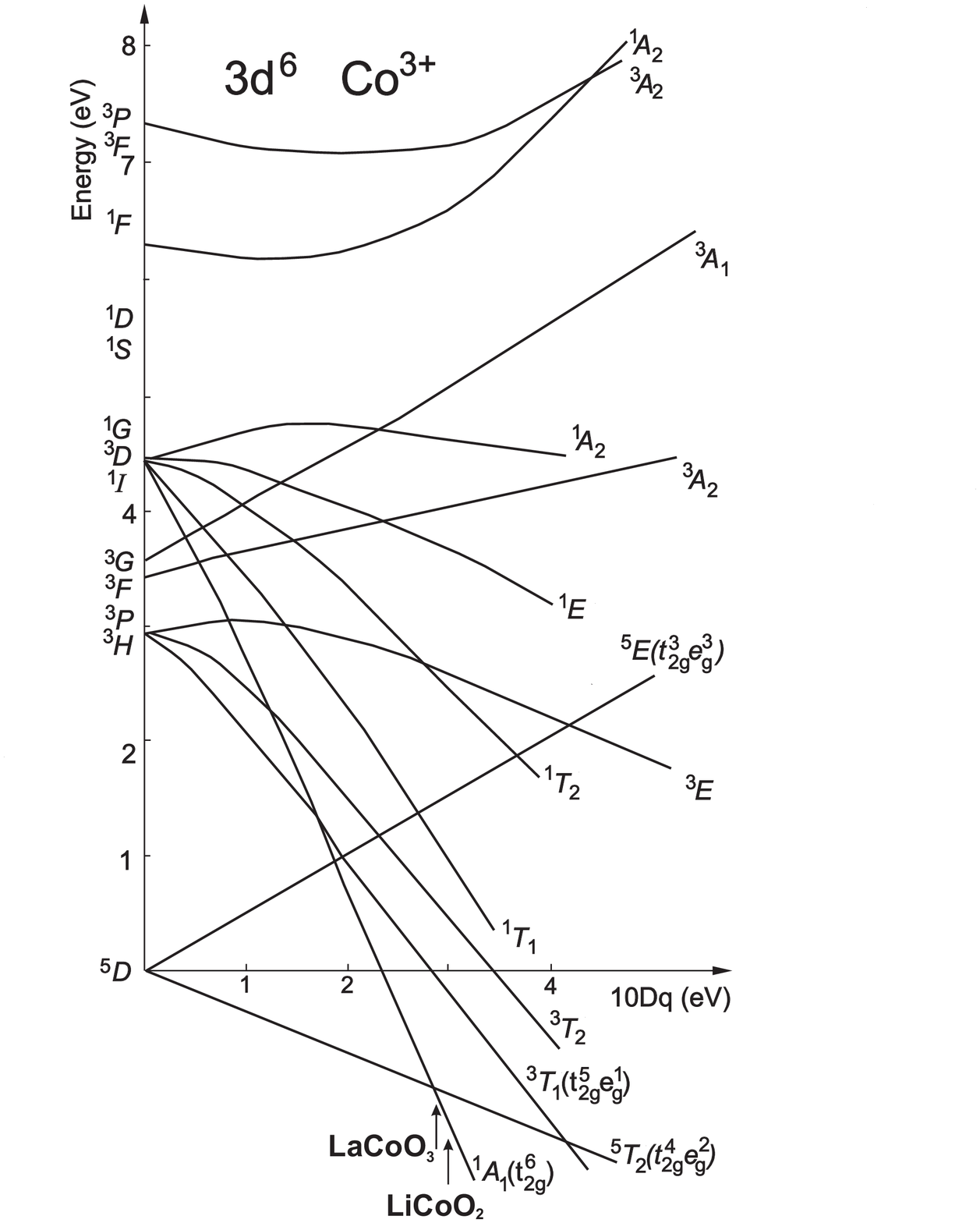}
\end{center}\vspace {-0.3 cm}
\caption{The modified Tanabe-Sugano diagram for the Co$^{3+}$ ion
(3$d^{6}$ configuration) showing the effect of the octahedral
crystal field on the electronic terms of the free Co$^{3+}$ ion.
The electronic structure of cubic subterms, corresponding to
$10Dq$ of 2.8 and 3.3 eV, relevant to LaCoO$_{3}$ and LiCoO$_{2}$
respectively, are marked by arrows. \vspace {-0.3 cm}}
\end{figure}

\begin{figure}\vspace {-0.2 cm}
\begin{center}
\includegraphics[width = 7.7 cm]{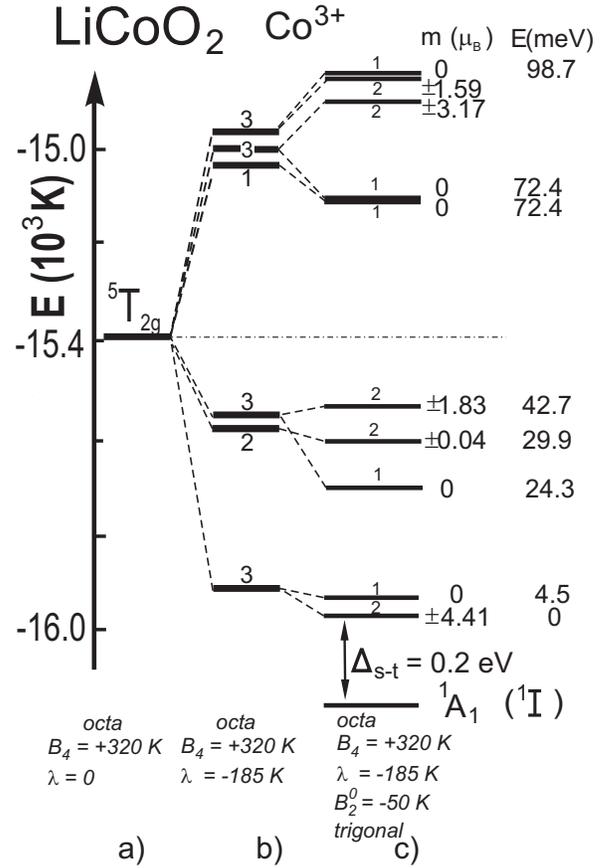}
\end{center} \vspace {-0.5 cm}
\caption{Calculated low-energy electronic structure of the
Co$^{3+}$ ion in
LiCoO$_{2}$ originating from the $^{5}T_{2g}$ cubic subterm with the $%
^{1}A_{1}$ singlet ground subterm put 0.2 eV below the lowest
$^{5}T_{2g}$ state. Such the structure is produced by the
dominant octahedral crystal-field interactions and the
intra-atomic spin-orbit coupling (b). c) shows the splitting
produced by the trigonal elongation distortion. The states are
labeled by the degeneracy, the magnetic moment and the energy
with respect to the lowest state of the $^{5}D$ term.}
\end{figure}

In stoichiometric LiCoO$_{2}$ there are Co$^{3+}$ ions only. The
Co$^{3+}$ ion has six electrons in the incomplete 3$d$ shell
forming, according to us, a strongly-correlated atomic-like
3$d^{6}$ configuration. It preserves its integrity also in a
solid. Such atomic-like system in a crystal experiences the
electrostatic crystal field (CEF) due to all surrounded charges.
This CEF modifies the term structure in the well-known and fully
controlled way, Fig. 1, and substantially removes the degeneracy
in the spin-orbital space, Fig. 2.\vspace {-0.5 cm}

\section{Results and discussion}\vspace {-0.3 cm}
The electronic structure of the Co$^{3+}$ ion is known from the
atomic physics and is given in the NIST database \cite{13}. 210
states are grouped in 16 atomic terms. The effect of the
octahedral crystal-field has been calculated already 50 years ago
by Tanabe and Sugano \cite{14,15,16}. A modified Tanabe-Sugano
diagram for the Co$^{3+}$ ion is shown in Fig. 1. For a
relatively weak octahedral CEF, $10Dq$ $<$ 2.75 eV, the ground
state is a state originating from the high-spin $^{5}T_{2g}$
cubic subterm - such a situation is realized in FeBr$_{2}$ and it
prefers a magnetic ground state \cite {12}. In LaCoO$_{3}$ the
octahedral CEF is stronger, $10Dq$ = 2.8 eV, due to the doubly
negative charge of the O ions compared to the single valency of
the Br ions and especially due to a small Co-O distance of 193
pm. As a consequence a low-spin $^{1}A_{1}$ ($^{1}I$) cubic
subterm (singlet!!) becomes the ground state \cite {2}. In
LiCoO$_{2}$ the Co-O distance is equally small like in
LaCoO$_{3}$ \cite{8} - note that in CoO the Co-O distance amounts
to 213 pm, being more than 10$\%$ larger. In case $10Dq$ $>$ 2.75
eV a non-magnetic state is formed in the atomic scale and such
the situation is realized in LiCoO$_{2}$.

0.2 eV above the $^{1}A_{1}$ non-magnetic state there is the
lowest state of the high-spin $^{5}T_{2g}$ cubic subterm (15-fold
degeneration) originating from the ionic 25-fold degenerated (in
the spin-orbit space) $^{5}D$ term. We have calculated its
splitting in a trigonal off-octahedral crystal field in the
presence of the spin-orbit coupling as is shown in Fig. 2. The
used negative sign of $B_{2}^{0}$ corresponds to the trigonal
elongation as is experimentally observed. Thus in the ionic model
we have got a consistent description of the insulating and
non-magnetic ground state. We used only three parameters:
octahedral crystal-field parameter $B_{4}^{z}$ = +320 K
$\Longleftrightarrow$ $10Dq$ = 3.3 eV); the spin-orbit coupling
$\lambda$ = -185 K and the trigonal distortion parameter
$B_{2}^{0}$ = -50 K. All these three parameters have clear
physical meaning and can be calculated really from {\it first
principles} \cite{17}. Of course, the most important assumption is
the strong-correlated approach preserving ionic integrity of the
3$d^{6}$ configuration.

We propagate the ionic model. We do not claim, of course, to
invent the crystal-field theory, because it was invented in
1929-1932 by Bethe, Van Vleck, Kramers, and many others, but in
last 20 years we have provided description of many compounds with
the use of the crystal-field theory with the importance of the
localized discrete states both in ionic and metallic systems. By
this we have proven applicability of the many-electron
crystal-field to the solid-state physics \cite{19}. We have
extended it from the single-ion theory to the Quantum Atomistic
Solid-State theory, QUASST, describing, for instance,
consistently both paramagnetic and magnetically-ordered states.
We have pointed out for 3$d$-ion compounds, for instance, the
importance of the spin-orbit coupling \cite{11} and of local
distortions. We assume in QUASST, in agreement with experiments,
on-site electron correlations to be sufficiently strong to keep
the atomic-like integrity of the 3$d$ ion in a solid.

We do not know reasons for forgetting or ignoring works on the
crystal field of early Van Vleck, Tanabe and Sugano, and of many
others. The Tanabe-Sugano diagrams have been almost forgotten in
the modern solid state theory \cite{19}, likely due to an
erroneous conviction that these states are not relevant to a
solid material. There is a discussion even on so simple problems
as description of the trigonal distortion used by us in Ref.
\cite{12} (it is correct) and calculated by us $^{5}E_{g}$
subterm as the ground state of the Mn$^{3+}$ ion in LaMnO$_{3}$.
An argument that the use of the crystal-field to a solid, in
particular to metallic magnetic rare-earth systems is erroneous
we consider as being not in agreement with experimnetal
observation. We take as very important the recent revealing in the
group of Prof. F. Steglich of localized CEF states in
UPd$_{2}$Al$_{3}$ and in heavy-fermion metal YbRh$_{2}$Si$_{2}$
\cite{20,21}. After my studies (RJR) with Prof. J. J. M. Franse
at the University of Amsterdam on metallic ErNi$_{5}$,
Ho$_{2}$Co$_{17}$, DyNi$_{5}$, Nd$_{2}$Fe$_{14}$B,
Pr$_{2}$Fe$_{14}$B all of these experimental facts prove,
according to us, a very substantial usefulness of the
crystal-field theory to a solid material, including metallic
magnetic rare-earth systems.

\begin{figure}\vspace {0 cm}
\begin{center}
\includegraphics[width = 8.5 cm]{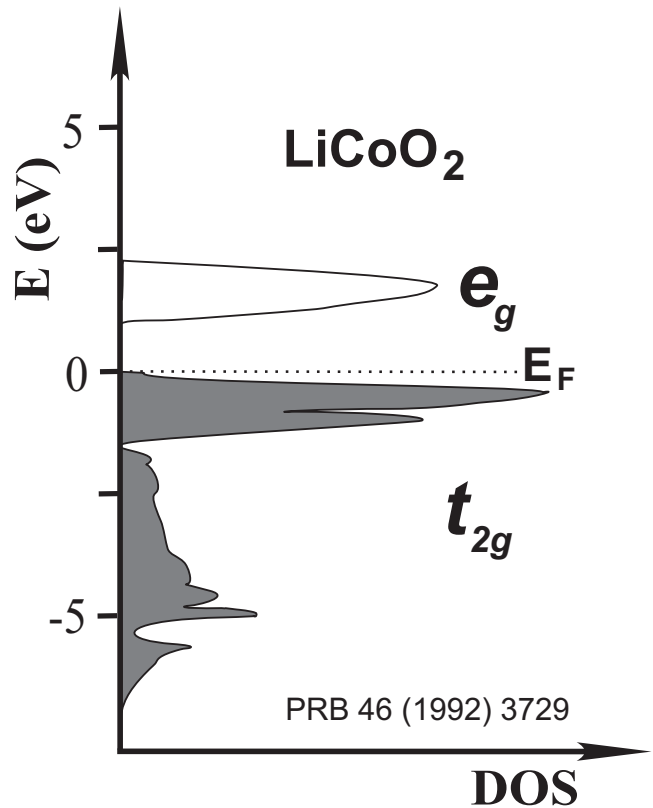}
\end{center} \vspace {-0.2 cm}
\caption{Schematic electronic structure of $d$ electrons in
LiCoO$_{2}$ obtained within the band theory, after Refs \cite{1}
(Fig.4) and \cite{8} (Fig.4a.) According to us this electronic
structure is not physically adequate.}
\end{figure}
One of reasons for the limited use of the Tanabe-Sugano diagrams
for 3$d$ compounds seems to be caused by the large uncertainty in
the strength of the octahedral CEF in a particular compound. From
this point of view the exact evaluation of the strength of the
octahedral CEF in LaCoO$_{3}$ for 2.8 eV \cite {2} we consider to
be of the great importance. Note that in the present literature
value of $10Dq$ varies in theoretical considerations of 3$d$
oxides from 0.5 eV \cite{6}, note 32 via 1.3 eV \cite{1} to a
poetical saying that it is very strong. QUASST seems to be a
standard approach to insulating transition-metal oxides but it has
been formulated in times of the ignorance of the crystal field
and the overwhelmed itinerant band treatment of 3$d$ electrons in
the modern solid-state theory. In Fig. 3 we present a band
treatment of $d$ electrons in LiCoO$_{2}$ from Refs \cite{1,8}
which we consider as not physically adequate. \vspace {-0.5 cm}
\section{Conclusions and remarks}
\vspace {-0.2 cm} We have described properties of LiCoO$_{2}$
within the Quantum Atomistic Solid State (QUASST) theory. It
starts for oxides from the purely ionic charge distribution and
takes into account very strong electron correlations,
predominantly of the intra-atomic origin, spin-orbit coupling and
the detailed local crystallographic surroundings and its
symmetry. Properties of LiCoO$_{2}$ are consistently explained
together with NaCoO$_{2}$ and LaCoO$_{3}$ - in all of these
compounds the Co$^{3+}$ ions occur in the low-spin state
($t_{2g}^{6}$, $S$ = 0) that is a crystal-field nonmagnetic
singlet $^{1}A_{1}$. This low-spin state is the effect of the
relatively strong crystal-field interactions ($B_{4}^{z}$ = +320
K $\Longleftrightarrow$ $10Dq$ = 3.3 eV) and is a manifestation
of the very large orbital moment of the Co$^{3+}$ ion. The
$^{1}A_{1}$ state originates from the atomic term $^{1}I$ that
has $L$ = 6. Thanks this large orbital moment the $^{1}A_{1}$
subterm becomes in sufficiently strong crystal field the ground
state breaking Hund's rules. We consider LiCoO$_{2}$ as a Mott
insulator basing on a definition that the Mott insulator occur
due to strong electron correlations despite of the open shell.

We claim that the many-electron strongly-correlated CEF approach
\cite {10}, the basis for QUASST, is physically adequate for 3$d$
oxides. The controversy about itinerant-localized treatment of
3$d$ electrons between one-electron CEF, many-electron CEF
(QUASST) approach and band approaches can be experimentally
solved by observation, or not, of the predicted electronic
structure. The comparison would be easier if band-structure
results contain data verifiable experimentally. At least,
effective charges of relevant atoms and experimental predictions
for zero-temperature properties and thermodynamics should be
reported. The QUASST calculations allow for {\it ab initio}
calculations of magnetic and electronic properties reconciling
for 3$d$ oxides both insulating, paramagnetic (LiCoO$_{2}$,
LaCoO$_{3}$, NaCoO$_{2}$) or magnetic (CoO, NiO, FeBr$_{2}$)
ground state.

Note: A paragraph, preceding the last one before the Conclusions,
with a discussion of the present scientific situation within the
magnetic community (13.07.2006) has been omitted.

$^\spadesuit$ dedicated to Hans Bethe and John H. Van Vleck,
pioneers of the crystal-field theory, to the 75$^{th}$
anniversary of the crystal-field theory, and to the Pope John
Paul II, a man of the freedom and the honesty in the human life
and in Science.

\end{document}